\documentclass[prl,amssymb,amsmath,twocolumn,groupedaddress,superscriptaddress,showpacs]{revtex4}

\usepackage{graphicx}
\usepackage{dcolumn}
\usepackage{bm}
\usepackage{epsfig}

\def\beaNN{\begin{eqnarray}\nonumber}
\def\bea{\begin{eqnarray}}
\def\eea{\end{eqnarray}}

\def\bef{\begin{figure}[htb]\begin{center}}
\def\befb{\begin{figure}[bh]\begin{center}}
\def\eef{\end{center}\end{figure}}

\def\mezzo{\frac{1}{2}}
\def\Np{n^+}
\def\Nm{n^-}
\def\Nz{n^0}
\def\Dg{D}
\def\At{A_{zz}^\mathrm{d}}

\def\reff#1{\ Fig.~(\ref{#1})}
\def\reqq#1{\ Eq.~(\ref{#1})}

\def\Ap{A_1^\mathrm{d}}
\def\Ad{A_1^\mathrm{d}}
\def\Fd{F_1^\mathrm{d}}
\def\Fdd{F_2^\mathrm{d}}
\def\Fdp{F_2^\mathrm{p}}
\def\Fdn{F_2^\mathrm{n}}

\def\A{{\cal A}}

\def\bd{b_1^\mathrm{d}}
\def\fd{F_1^\mathrm{d}}
\def\gd{g_1^\mathrm{d}}
\def\gn{g_1^\mathrm{n}}

\def\dsigma#1{\frac{\mathrm{d}^2#1}{\mathrm{d}x\mathrm{d}Q^2}}
\def\VV{P_z}
\def\TT{P_{zz}}
\def\TTe{P_{zz}^{\rm eff}}
\def\su{\sigma_U}

\def\jum#1{\hspace*{#1cm}}
\def\vum#1{\vspace*{#1cm}}

\def\cita#1{\cite{#1}}
\def\reff#1{Fig.~\ref{#1}}
\def\ret#1{Table~\ref{#1}}
\def\req#1{Eq.~(\ref{#1})}

\def\cT{c_{zz}}

\def\Journal#1#2#3#4{{#1} {\bf #2}, #3 (#4).}

\def\NIMA{{ Nucl. Instr. Meth.} A}

\def\NPB{{ Nucl. Phys.} B}
\def\PLB{{ Phys. Lett.}  B}
\def\PRL{ Phys. Rev. Lett.}
\def\PRD{{ Phys. Rev.} D}
\def\PRC{{ Phys. Rev.} C}

\def\ZPA{{ Z. Phys.} A}

\def\PR{{ Phys. Rev.} }
\def\NPA{{ Nucl. Phys.}  A}
\def\ZPA{{ Z. Phys.}  A}

\def\EPJA{{ Eur. Phys. J. A} }
\def\SOPD{{ Sov. Phys. Doklady} }
\def\SJNP{{ Sov. J. Nucl. Phys.} }

\def\Hermes{{\sc Hermes\ }}

\def\HERA{{\sc Hera\ }}
\def\Hera{{\sc Hera}}
\def\DESY{{\sc Desy\ }}
\def\EMC{{\sc EMC}}
\def\NMC{{\sc NMC}}

\def\sspar{\sigma^{\stackrel{\rightarrow}{\Rightarrow}}}
\def\ssant{\sigma^{\stackrel{\rightarrow}{\Leftarrow}}}
\def\sspiu{\sigma^{\Leftrightarrow}}
\def\ssnul{\sigma^{{0}}}
\def\polpar{P_{zz}^{\stackrel{\rightarrow}{\Rightarrow}}}
\def\polant{P_{zz}^{\stackrel{\rightarrow}{\Leftarrow}}}
\def\polpiu{P_{zz}^{\Leftrightarrow}}
\def\polnul{P_{zz}^{{\;0}}}
\def\ssuno{\sigma^{{1}}}

\def\quup{q^1_{\uparrow}}
\def\qudp{q^{1}_{\downarrow}}
\def\quum{q^{-1}_{\uparrow}}

\def\quuz{q^0_{\uparrow}}
\def\qudz{q^0_{\downarrow}}
\def\qup{q^\mezzo_{\uparrow}}
\def\qum{q^{-\mezzo}_{\uparrow}}
\def\qdp{q^\mezzo_{\downarrow}}

\def\heliz{helicity-$0$}
\def\Heliz{helicity-$0$\ }

\begin{document}


\title{First Measurement of the Tensor Structure Function $b_1$ of the Deuteron.}




\def\groupalberta{\affiliation{Department of Physics, University of Alberta, Edmonton, Alberta T6G 2J1, Canada}}
\def\groupargonne{\affiliation{Physics Division, Argonne National Laboratory, Argonne, Illinois 60439-4843, USA}}
\def\groupbari{\affiliation{Istituto Nazionale di Fisica Nucleare, Sezione di Bari, 70124 Bari, Italy}}
\def\groupbeijing{\affiliation{School of Physics, Peking University, Beijing 100871, China}}
\def\groupchina{\affiliation{Department of Modern Physics, University of Science and Technology of China, Hefei, Anhui 230026, China}}
\def\groupcolorado{\affiliation{Nuclear Physics Laboratory, University of Colorado, Boulder, Colorado 80309-0390, USA}}
\def\groupdesy{\affiliation{DESY, 22603 Hamburg, Germany}}
\def\groupzeuthen{\affiliation{DESY, 15738 Zeuthen, Germany}}
\def\groupdubna{\affiliation{Joint Institute for Nuclear Research, 141980 Dubna, Russia}}
\def\grouperlangen{\affiliation{Physikalisches Institut, Universit\"at Erlangen-N\"urnberg, 91058 Erlangen, Germany}}
\def\groupferrara{\affiliation{Istituto Nazionale di Fisica Nucleare, Sezione di Ferrara and Dipartimento di Fisica, Universit\`a di Ferrara, 44100 Ferrara, Italy}}
\def\groupfrascati{\affiliation{Istituto Nazionale di Fisica Nucleare, Laboratori Nazionali di Frascati, 00044 Frascati, Italy}}
\def\groupgent{\affiliation{Department of Subatomic and Radiation Physics, University of Gent, 9000 Gent, Belgium}}
\def\groupgiessen{\affiliation{Physikalisches Institut, Universit\"at Gie{\ss}en, 35392 Gie{\ss}en, Germany}}
\def\groupglasgow{\affiliation{Department of Physics and Astronomy, University of Glasgow, Glasgow G12 8QQ, United Kingdom}}
\def\groupillinois{\affiliation{Department of Physics, University of Illinois, Urbana, Illinois 61801-3080, USA}}
\def\groupmit{\affiliation{Laboratory for Nuclear Science, Massachusetts Institute of Technology, Cambridge, Massachusetts 02139, USA}}
\def\groupmichigan{\affiliation{Randall Laboratory of Physics, University of Michigan, Ann Arbor, Michigan 48109-1040, USA }}
\def\groupmoscow{\affiliation{Lebedev Physical Institute, 117924 Moscow, Russia}}
\def\groupmunich{\affiliation{Sektion Physik, Universit\"at M\"unchen, 85748 Garching, Germany}}
\def\groupnikhef{\affiliation{Nationaal Instituut voor Kernfysica en Hoge-Energiefysica (NIKHEF), 1009 DB Amsterdam, The Netherlands}}
\def\groupstpetersburg{\affiliation{Petersburg Nuclear Physics Institute, St. Petersburg, Gatchina, 188350 Russia}}
\def\groupprotvino{\affiliation{Institute for High Energy Physics, Protvino, Moscow region, 142281 Russia}}
\def\groupregensburg{\affiliation{Institut f\"ur Theoretische Physik, Universit\"at Regensburg, 93040 Regensburg, Germany}}
\def\grouprome{\affiliation{Istituto Nazionale di Fisica Nucleare, Sezione Roma 1, Gruppo Sanit\`a and Physics Laboratory, Istituto Superiore di Sanit\`a, 00161 Roma, Italy}}
\def\groupsimonfraser{\affiliation{Department of Physics, Simon Fraser University, Burnaby, British Columbia V5A 1S6, Canada}}
\def\grouptriumf{\affiliation{TRIUMF, Vancouver, British Columbia V6T 2A3, Canada}}
\def\grouptokyo{\affiliation{Department of Physics, Tokyo Institute of Technology, Tokyo 152, Japan}}
\def\groupamsterdam{\affiliation{Department of Physics and Astronomy, Vrije Universiteit, 1081 HV Amsterdam, The Netherlands}}
\def\groupwarsaw{\affiliation{Andrzej Soltan Institute for Nuclear Studies, 00-689 Warsaw, Poland}}
\def\groupyerevan{\affiliation{Yerevan Physics Institute, 375036 Yerevan, Armenia}}
\def\groupnone{\noaffiliation}


\groupalberta
\groupargonne
\groupbari
\groupbeijing
\groupchina
\groupcolorado
\groupdesy
\groupzeuthen
\groupdubna
\grouperlangen
\groupferrara
\groupfrascati
\groupgent
\groupgiessen
\groupglasgow
\groupillinois
\groupmit
\groupmichigan
\groupmoscow
\groupmunich
\groupnikhef
\groupstpetersburg
\groupprotvino
\groupregensburg
\grouprome
\groupsimonfraser
\grouptriumf
\grouptokyo
\groupamsterdam
\groupwarsaw
\groupyerevan


\author{A.~Airapetian}  \groupyerevan
\author{N.~Akopov}  \groupyerevan
\author{Z.~Akopov}  \groupyerevan
\author{M.~Amarian}  \groupzeuthen \groupyerevan
\author{V.V.~Ammosov}  \groupprotvino
\author{A.~Andrus}  \groupillinois
\author{E.C.~Aschenauer}  \groupzeuthen
\author{W.~Augustyniak}  \groupwarsaw
\author{R.~Avakian}  \groupyerevan
\author{A.~Avetissian}  \groupyerevan
% ...


\collaboration{The HERMES Collaboration}
\noaffiliation

\date{\today}

\begin{abstract}
The \Hermes experiment has investigated the tensor spin structure of the deuteron using the $27.6$~GeV/c
positron beam of \Hera.  The use of a tensor polarized deuteron gas target with only a 
negligible residual vector polarization enabled the first measurement of the tensor asymmetry $\At$ and the tensor structure 
function $\bd$ for average values of the Bj\o rken variable $0.01\!<\!\langle x\rangle\!<\!0.45$ and of the negative 
of the squared four-momentum transfer $0.5\; {\rm GeV^2}\!<\!\langle Q^2 \rangle \!<\!5\;{\rm GeV^2}$. 
The quantities $\At$ and $\bd$ are found to be 
non-zero. The rise of $\bd$ for decreasing values of $x$ can be interpreted 
to originate from the same mechanism that leads to nuclear shadowing in unpolarized scattering.
\end{abstract}


\maketitle


Inclusive deep-inelastic scattering (DIS) of charged leptons from the deuteron, a spin-$1$ object, 
is described by eight structure functions, twice as many as required to describe DIS from a  
spin-$1/2$ nucleon~\cita{jaffe}. 
The three leading-twist structure functions relevant to this discussion can be written within
the Quark-Parton Model as: 

\vum{0.2}
\noindent
\begin{tabular}{ccc}  
\vum{0.2}  & Nucleon & Deuteron \\

\vum{0.2} $ F_1$ \jum{0.2} & $ \mezzo \sum_q e_q^2 \,[\qup+\qum]$ \jum{0.2} & $ \frac{1}{3}\sum_q e_q^2 \,[\quup + \quum +\quuz]$\\
\vum{0.2} $ g_1$ \jum{0.2} & $ \mezzo\sum_q e_q^2 \,[\qup-\qdp]$  \jum{0.2} & $ \mezzo\sum_q e_q^2 \,[\quup-\qudp]$  \\
\vum{0.2} $ b_1$ \jum{0.2} & $--$ & $\mezzo \sum_q e_q^2 \left[2\quuz-(\quup+\quum)\right]$ .
\end{tabular}


\noindent
where $q^{m}_{\uparrow}$($q^{m}_{\downarrow}$) is the number density of 
quarks with spin up(down) along the $z$ axis in a hadron(nucleus) with
helicity $m$ moving with infinite momentum along the $z$ axis.  
Reflection symmetry implies that $q^{m}_{\uparrow}=q^{-m}_{\downarrow}$.
The sums run over quark and
antiquark flavors $q$ with a charge $e_q$ in units of the elementary charge $e$.
Both structure functions and quark number densities depend on the Bj\o rken variable $x$
which can be interpreted as the fraction of the nucleon momentum carried by the 
struck quark in the infinite-momentum frame, and $-Q^2$, the square of the four-momentum 
transfer by the virtual photon.  

The polarization-averaged structure function $F_1$ describes the quark distributions averaged
over the target spin states. The polarization-dependent structure function $g_1$
describes the imbalance in the distribution of quarks with the same $q^m_{\uparrow}$ or 
opposite $q^m_{\downarrow}$ helicity with respect to that of the parent hadron. 
It can be measured only when both beam and target are polarized. 
The tensor structure function $b_1$ does not exist for spin-$1/2$ targets and vanishes in the absence of 
nuclear effects, i.e.~if the deuteron simply consists of a proton and neutron in a relative $S$-state. 
It describes the difference in the 
quark distributions between the \heliz, $q^0\!=\!(\quuz+\qudz)\!=\!2\quuz$, and the averaged non-zero helicity, 
$q^1\!=\!(\quup+\qudp)\!=\!(\quup + \quum)$, states of the deuteron~\cite{jaffe,pais,frankfurt}. 
Because $b_1$ depends only on the spin averaged quark distributions 
$b_1=\mezzo \sum_q e_q^2 \left[q^0-q^1\right]$,
its measurement does not require a polarized beam.  
Since the magnitude of $\bd$ was expected to be small, 
it was usually ignored in the extraction of 
$\gd$ and, as a consequence, of
the neutron structure function $\gn$ derived from deuteron and proton data. 
This is in general not {\it a priori} justified. 
This paper reports the first measurement of $\bd$, performed by the \Hermes 
collaboration using a data set taken with a positron beam and a tensor-polarized deuterium target, and an
integrated luminosity of 42~$\rm pb^{-1}$.

The \Hermes experiment~\cita{hermes} was designed to investigate the internal spin structure of nucleons and
nuclei by deep-inelastic scattering of polarized positrons and electrons by polarized gaseous targets (e.g.~hydrogen, 
deuterium and helium--3) internal to the \Hera-$e$ storage ring. The positrons (electrons) circulating in the ring 
become transversely polarized by the emission of spin-flip synchrotron radiation~\cita{sokolov}. 
A longitudinal beam polarization is generated at \Hermes by the 
use of a pair of spin rotators before and after the experiment.
Beam polarization is employed for the simultaneous 
$\gd$ measurement and for studies on beam related systematic effects on $\bd$. 


A feature of \Hermes unique among polarized DIS experiments is its atomic-gas target that is not diluted by 
non-polarizable material~\cita{ABS}. The target cell is 
an open-ended elliptical aluminum tube internal to the beam line (40~cm long, $75$~$\mu \rm m$ 
in wall thickness), 
which is used to confine the polarized gas along the beam. A magnetic field 
surrounding the target and aligned with the beam provides the quantization axis for the nuclear spin. 
A sample of gas was continuously drawn from the cell 
and analyzed to determine the atomic and
molecular abundances and the nuclear polarization of the atoms. 
An atomic beam source (ABS) generates a deuterium atomic beam and selects the 
two hyperfine states with the desired nuclear polarization to be injected into the cell.
This system allows the selection of substates with pure tensor polarization 
$\TT\,=\,(\Np\!+\!\Nm - 2\Nz)/(\Np\!+\!\Nm\!+\!\Nz)$
and at the same time vanishing vector polarization 
$ \VV\,=\,(\Np\!-\!\Nm)/(\Np\!+\!\Nm\!+\!\Nz)$, 
a combination which is not possible for solid-state polarized targets.
%
%
Here $\Np,\,\Nm,\,\Nz$ are the atomic populations with positive, negative and zero spin projection on 
the beam direction, respectively. 
Every $90$~seconds the polarization of the injected gas is changed; for the $\bd$ measurement the four injection 
modes listed in \ret{Tabcross} were continuously cycled.
Note that the vector$^+$ and vector$^-$ modes are also employed for the $\gd$ measurement.


\begin{table}[b]
\caption{ \label{Tabcross}
Hyperfine state composition, corresponding atomic population and
polarization of the target states~\cita{ABS} employed in the $\bd$ measurement as described in the text.
The average target vector $\VV$ and tensor $\TT$ polarizations are typically more than $80\%$ of the 
ideal values. 
The average beam polarization $|P_B|$ is $0.53\pm0.01$. A positive $P_B$ is assumed in the table.
Four independent polarized yields, defined in the text, were measured depending on beam and 
target polarizations.} 
\vum{-0.5}
\begin{center}
\begin{tabular}{ccccccc} \hline\hline
Target  & Hyper. & Atomic & Tensor term   & Vector term         & Meas.     \\
state  & state  & popul. &    $\TT$       & $\VV\cdot P_B$      & yield     \\
\hline
vector$^+$ & $|1\rangle\!+\!|6\rangle$ & $n^+$         & $+0.80\pm0.03$  & $+0.45\pm0.02$  & $\sspar$     \\
vector$^-$ & $|3\rangle\!+\!|4\rangle$ & $n^-$         & $+0.85\pm0.03$  & $-0.45\pm0.02$  & $\ssant$   \\
tensor$^+$ & $|3\rangle\!+\!|6\rangle$ & \jum{.05} $n^+\!+\!n^-$ & $+0.89\pm0.03$  & \jum{0.20} $0.00\pm0.01$  &  $\sspiu$ \\ \hline
tensor$^-$ & $|2\rangle\!+\!|5\rangle$ & $n^0$         & $-1.65\pm0.05$  & \jum{0.20} $0.00\pm0.01$  &  $\ssnul$\\
\hline\hline
%
\end{tabular}
\end{center}
\vum{-0.4}
\end{table}

The \Hermes detector~\cita{hermes} is a forward spectrometer with a dipole magnet providing a field integral of
1.3 Tm. A horizontal iron plate shields the \HERA beam lines from this field, thus dividing the 
spectrometer into two identical halves with $\pm 170$ mrad horizontal and
$\pm 40$ to $\pm 140$ mrad vertical acceptance for the polar scattering angle. 
Tracking is based on $36$ drift chamber planes in each detector-half.
Positron identification is accomplished using a likelihood method based on signals of four subsystems: a ring-imaging 
$\rm \Check{C}erenkov$ detector, a lead-glass calorimeter, a transition-radiation detector, and a preshower 
hodoscope. For positrons in the momentum range of $2.5$ GeV/c to $27$ GeV/c, the identification efficiency 
exceeds $98\%$ and the hadron contamination is less than $0.5 \%$. The average polar angle resolution is $0.6$ 
mrad and the average momentum resolution is $2 \%$. 

For a target state characterized by $\VV$ and $\TT$, the DIS yield measured by the experiment is 
proportional to the double differential cross section of polarized DIS
\bea
\dsigma{\sigma_P} &\simeq & \dsigma{\sigma} \left[1 \,-\, \VV P_B\Dg \Ap \,+\, \mezzo \TT \At\right]. \jum{0.4}
\label{CrossEq}
\eea
Here, $\sigma$ is the unpolarized cross section, 
$\Ap$ is the vector and $\At$ the tensor asymmetry of the virtual-photon deuteron cross section, $P_B$ is the 
beam polarization and $\Dg$ is the fraction of the beam polarization transferred to the virtual photon. 
In \req{CrossEq} and in the following \req{Eqgmeas}, the fractional correction ($\lesssim$ 0.01) arising from 
the interference between longitudinal and transverse photo-absorption 
amplitudes, which leads to the structure function $g_2$~\cita{slacg2}, is neglected. 
Four independent polarized yields were measured (see \ret{Tabcross} for the values of the achieved polarizations): 
$\sspar$ and $\ssant$ when the target spin is parallel and anti-parallel to that of the beam, respectively, 
$\sspiu$ when the target has a mixture of helicity-1 states, and $\ssnul$ when the target is in the \Heliz state.
The tensor asymmetry is extracted as
%
\bea
\At &=& \frac{2\ssuno \,-\, 2\ssnul}{3\su \TTe},
\label{EqAzz}
\eea
where $\ssuno=(\ssant + \sspar + \sspiu) / 3$ is the average over the helicity-1 states, 
$\su=(2\ssuno + \ssnul)/3$ is the polarization-averaged yield and $\TTe=(\polpar+\polant+\polpiu-3\polnul)/9$ 
is the effective tensor polarization. In \req{EqAzz} the vector component due to $\Ap$ nearly cancels out:
a residual vector polarization not more than 0.02 is achieved both for the $\ssnul$, $\sspiu$ and the 
averaged $(\ssant+\sspar)/2$ measurements,
(see \ret{Tabcross}). Note that any contribution from residual vector polarization of the target is reduced to a 
negligible level by grouping together two sets of data with approximately the same statistics and opposite beam 
helicities. 

The polarization-dependent structure function $\gd$ can be extracted from the vector asymmetry $\Ap$ as
%
\bea
\frac{\gd}{\Fd} \,\simeq\, \Ad &\simeq& 
\frac{\cT}{|\VV P_B|D}\,\frac{(\ssant\,-\,\sspar)}{(\ssant\,+\,\sspar)}, 
\label{Eqgmeas} 
\eea
\bea
{\rm with} \jum{0.2} \cT \!&=& \frac{(\ssant\,+\,\sspar)}{2\su}\,=\,1+\frac{(\polant+\polpar)}{4}\At.  \jum{0.4}
\label{EqTbias}
\eea
In all previous determinations of $\gd$ the contribution of the tensor asymmetry $\At$ was neglected,
i.e.~$\cT$ was assumed to be equal to $1$, in spite of the fact that the vector polarization of the target 
could only be generated together with a non-zero tensor polarization. The present measurement 
quantifies the effect of $\At$ on the existing $\gd$ data.
 
For the determination of $\At$, about 3.2 million inclusive events obtained with a tensor-polarized deuterium 
target are selected,
by requiring as in the \Hermes $\gd$ analysis \cita{g1d} a scattered positron with $0.1\;{\rm GeV^2}<Q^2<20\;{\rm GeV^2}$ 
and an invariant mass of the virtual-photon nucleon system $W>1.8$ GeV.
The kinematic range covered by the selected data is 
$0.002 < x < 0.85$ and $0.1 < y < 0.91$, where $y$ is the fraction of the beam energy carried by the virtual 
photon in the target rest frame.  
The asymmetry $\At$ is calculated according to \req{EqAzz}.
%
The number of events determined per spin state is corrected for the $e^+ e^-$ background arising 
from charge symmetric processes (the latter is negligible at high $x$ but amounts to almost $15\%$
of the statistics in the lowest-$x$ bin) and normalized to the luminosity measured by Bhabha 
scattering from the target gas electrons~\cita{hermes}. 

\begin{figure}[t]
\includegraphics[width=0.70\linewidth]{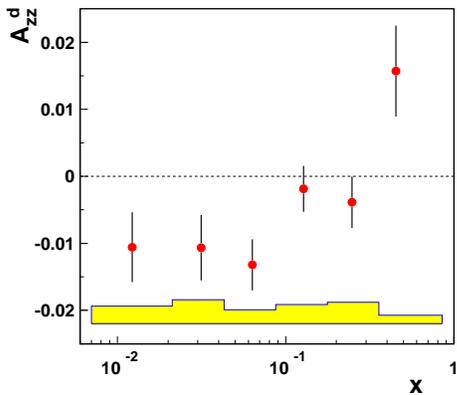}
\caption{\small The tensor asymmetry $\At(x)$. 
The error bars are statistical and the shaded band shows the systematic uncertainty. \vum{-0.5}}
\label{Figuno}
\end{figure}

The asymmetry $\At$ is corrected for detector smearing and QED radiative effects to obtain the Born 
asymmetry corresponding to pure single-photon exchange in the scattering process. 
The kinematic 
migration of the events due to radiative and detector smearing is treated using an unfolding algorithm, 
which is only sensitive to the detector model, the known unpolarized cross section, and the 
models for the background processes~\cita{dqlong}. 
The radiative background is negligible at high $x$ but increases as $x\rightarrow 0$ and 
reaches almost 50 \% of the statistics in the lowest-$x$ bin. 
The radiative corrections 
are calculated using a Monte Carlo generator based on RADGEN~\cita{radgen}. 
The coherent and quasi-elastic radiative 
tails are estimated using parameterizations of the deuteron form factors~\cita{elas_new,qela_new} and
corrected for the tracking inefficiency due to showering of the radiated photons.
The polarized part of the quasi-elastic radiative tail is neglected since there is no net tensor 
effect by inclusive scattering on weakly-bound spin-$1/2$ objects~\cita{bulten}.
The extracted tensor asymmetry $\At$ is shown in \reff{Figuno} and listed in \ret{b1result}. 
It appears to be positive at high $x$ and is negative at low $x$, crossing zero at an $x$ value of about $0.2$.
In the lowest-$x$ bin, the asymmetry is not zero 
at the 2-sigma level only after the subtraction of the radiative background.
The magnitude of the observed tensor asymmetry $\At$ does not exceed~$0.02$ over the measured range; from this 
result and using Eqs.~(\ref{Eqgmeas}) and (\ref{EqTbias}), the fractional correction to the \Hermes $\gd$ measurement 
due to the tensor asymmetry is estimated to be less than $0.01$. 

The particle identification efficiency and the target polarization measurement
give negligible contributions to the systematic uncertainty. 
The normalization 
uncertainty between different injection modes of the ABS is $\approx 1\times 10^{-3}$ 
and correlated over the kinematic bins. This uncertainty is estimated by the observed 2-sigma 
offset from zero of the asymmetry between averaged vector, $2\ssuno=\sspar+\ssant$, 
and tensor$^+$, $\ssnul$ replaced by $\sspiu$ in \req{EqAzz}, non-zero helicity injection modes. 
The luminosity 
measurement is sensitive to possible residual polarization of the target gas electrons.
The asymmetries obtained by normalizing the yields to the luminosity-monitor rates, or to the 
beam-current times the target-gas analyzer rates, are in good agreement within 
the quoted normalization uncertainty.
The subtraction of the radiative background inflates the size of the statistical and
the above mentioned systematic uncertainties by almost a factor of 2 at low $x$.
The systematic uncertainty of the radiative corrections is 
$\approx 2\times 10^{-3}$ for the three bins at low $x$ and negligible at high $x$. 
A possible misalignment in the spectrometer geometry yields an uncertainty $\approx 3\times 10^{-3}$
in the bins where the asymmetry changes sign.
All the contributions to the systematic uncertainty are added in quadrature.
The two subsamples of data with opposite beam helicities were analyzed independently and gave 
consistent $\At$ results. 

\begin{figure}[t]
\includegraphics[width=0.75\linewidth]{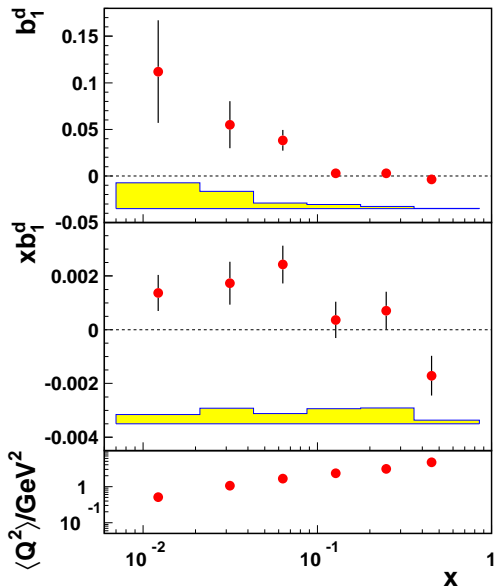}
\caption{\small The tensor structure function presented as (top) $\bd(x)$ and (middle) $x\bd(x)$. 
The error bars are statistical and the shaded bands show the systematic uncertainty. The bottom 
panel shows the average value of $Q^2$ in each $x$-bin. \vum{-0.5}}
\label{Figdue}
\end{figure}

The tensor structure function $\bd$ is extracted from the tensor asymmetry using the relations~\cite{weise,ssratio}
\bea
\bd\,=\, -\frac{3}{2}\At \, \Fd &;\jum{1} & \Fd\,=\,\frac{(1+Q^2/\nu^2) \Fdd }{2x(1+R)}.
\label{Eqextraction}
\eea
No contribution from the hitherto unmeasured double spin-flip structure function $\Delta$~\cita{gluon} is 
considered here, being kinematically suppressed for a longitudinally polarized target~\cita{sather}.
The structure function $\Fdd$ is calculated as $\Fdd\,=\,\Fdp (1+\Fdn/\Fdp)/2$ using the parameterizations
of the precisely measured structure function $\Fdp$~\cita{allm97} and $\Fdn/\Fdp$ ratio~\cita{ffratio}.
In \reqq{Eqextraction}, $R=\sigma_L/\sigma_T$ is the ratio of longitudinal to transverse photo-absorption cross 
sections~\cita{ssratio} and 
$\nu$ is the virtual-photon energy.
The results for $\bd$ are listed together with those for $\At$ in \ret{b1result}. 
The $x$-dependence of $\bd$ is displayed in \reff{Figdue}.
The data show that $\bd$ is different from zero for $x<0.1$, its magnitude rises for 
decreasing values of $x$ and, for 
$x \lesssim 0.03$, becomes even larger than that of $\gd$ at the same value of $Q^2$~\cita{g1d}. 


%
\begin{table}[b]
\caption{ \label{b1result}
Measured values (in $10^{-2}$ units) of the tensor asymmetry $\At$ and the tensor structure function $\bd$. 
Both the corresponding statistical and systematic uncertainties are listed as well.}
\vum{-0.0}
\begin{tabular}{ccrrrrrr} \hline\hline
$\langle x \rangle$ & $\langle Q^2\rangle$ & $\At$ \jum{0.1} & \jum{-0.3} $\pm\delta \A_{zz}^{\rm stat}$ & 
\jum{-0.2} $\pm\delta \A_{zz}^{\rm sys}$ &
\jum{0.3} $\bd$  \jum{0.1} & \jum{-0.3} $\pm\delta {b_1}^{\!\rm stat}$ & \jum{-0.2} $\pm\delta {b_1}^{\!\rm sys}$ \\
   & $\rm [GeV^2]$ & \jum{0.1} $[10^{-2}]$ & $[10^{-2}]$ & $[10^{-2}]$ & \jum{0.1} $[10^{-2}]$ & $[10^{-2}]$ & $[10^{-2}]$ \\
\hline
  0.012 &  0.51 &  -1.06 &  0.52 &   0.26 &  11.20 &   5.51 &   2.77 \\
  0.032 &  1.06 &  -1.07 &  0.49 &   0.36 &   5.50 &   2.53 &   1.84 \\
  0.063 &  1.65 &  -1.32 &  0.38 &   0.21 &   3.82 &   1.11 &   0.60 \\
  0.128 &  2.33 &  -0.19 &  0.34 &   0.29 &   0.29 &   0.53 &   0.44 \\
  0.248 &  3.11 &  -0.39 &  0.39 &   0.32 &   0.29 &   0.28 &   0.24 \\
  0.452 &  4.69 &   1.57 &  0.68 &   0.13 &  -0.38 &   0.16 &   0.03 \\ \hline\hline
\end{tabular}
\vum{-0.0}
\end{table}

Because the deuteron is a weakly-bound state of spin-$1/2$ nucleons, $\bd$ was initially
predicted to be negligible, at least at moderate and large values of $x$ ($x>0.2$)~\cita{manohar,umnikov},
where it should be driven by nuclear binding and Fermi motion effects. 
It was later realized that $\bd$ could rise to values which significantly differ from zero as 
$x\rightarrow 0$, and 
its magnitude could reach about $1\%$ of the unpolarized structure function $\fd$, due to
the same mechanism that leads to the well known effect of nuclear shadowing in unpolarized 
scattering~\cita{EMC}. This feature is described by coherent double-scattering 
models~\cita{miller,khan,strikman,nikolaev,edelmann,weise,bora}.
The observed $\bd$ confirms qualitatively the double-scattering model predictions, except for the 
negative value at $\langle x\rangle=0.452$, which, however, is still compatible with zero at the 
2-sigma level. In~the~context of the Quark-Parton Model description, the sum rule $\int b_1(x) \mathrm{dx} = 0$ is 
broken if the quark sea is tensor polarized~\cita{close, efremov}. From the $x$-behavior of $x\bd$ shown in \reff{Figdue} it
can be seen that the first moment of $\bd$ is non-zero. A 2-sigma result, $\int_{0.002}^{0.85} b_1(x) \mathrm{d}x = 
(1.05\pm 0.34_{\ \rm stat}\pm0.35_{\ \rm sys})\cdot 10^{-2}$, is obtained within the measured range, and a 1.7-sigma 
result, $\int_{0.02}^{0.85} b_1(x) \mathrm{d}x = (0.35\pm 0.10_{\ \rm stat}\pm0.18_{\ \rm sys})\cdot 10^{-2}$, 
within the restricted $x$-range where $Q^2\!>\!1$ $\rm GeV^2$.
The integrals are calculated after having $\bd$ evolved to
$Q_0^2=5$ $\rm GeV^2$ by assuming a $Q^2$-independence of the measured $\bd/\Fd$ ratio,
$\bd(Q_0^2)=\bd/\Fd\cdot\Fd(Q_0^2)$.

In conclusion, \Hermes has provided the first measurement of the tensor structure function $\bd$, 
in the kinematic domain $0.01\!<\!\langle x\rangle\!<\!0.45$ and $0.5\;{\rm GeV^2}\!<\!\langle Q^2 \rangle \!<\!5\;{\rm GeV^2}$. 
The function $\bd$ is found to be different from zero for $x<0.1$. Its first moment is found to be 
not zero at the 2-sigma level within the measured $x$ range. 
The $\bd$ measurement can be used to reduce the systematic uncertainty on the $\gd$ 
measurement that is assigned to the tensor structure of the deuteron. The behavior of $\bd$ at 
low values of $x$ is in qualitative agreement with expectations based on coherent double-scattering models.

We gratefully acknowledge the \DESY management for its support, the staff at \DESY and the
collaborating institutions for their significant effort, and our national funding agencies for
financial support. We are grateful to N. Nikolaev and O. Shekhovtsova for useful discussions.
\bibliography{basename of .bib file}

\end{document}